\documentclass[floatfix,lengthcheck,showpacs,amssymb,amsmath,amsfonts,twocolumn]{aastex63}
\usepackage{CJK}
\usepackage[utf8]{inputenc}
\usepackage{color}
\usepackage{graphicx}

\usepackage{float}
\usepackage{subfigure}
\usepackage{physics}
\usepackage{amsmath}
\usepackage{afterpage}
\usepackage{acronym}
\usepackage{multirow}
\usepackage{tabularx}
\usepackage{booktabs}
\usepackage{tikz}
\usetikzlibrary{arrows,shapes,trees,decorations.pathreplacing}
\usepackage{import}
\usepackage{svn-multi}
\usepackage{hyperref}
\usepackage{gensymb}
\usepackage{bm}
\usepackage{acronym}
\hypersetup{
    colorlinks=true,
    linkcolor=blue,
    filecolor=magenta,      
    urlcolor=cyan,
}

\newcommand{\AEI}{\affiliation{Max-Planck-Institut f{\"u}r Gravitationsphysik (Albert-Einstein-Institut), D-30167 Hannover, Germany}}
\newcommand{\UniHannover}{\affiliation{Leibniz Universit{\"a}t Hannover, D-30167 Hannover, Germany}}

\begin{document}
    \newacro{nr}[NR]{Numerical relativity}
    
    \newacro{gw}[GW]{gravitational-wave}

    \newacro{gws}[GWs]{gravitational waves}
    
    \newacro{pe}[PE]{parameter estimation}

    \newacro{psd}[PSD]{Power Spectral Density}

    \newacro{em}[EM]{electromagnetic}

\begin{CJK*}{UTF8}{gbsn}
\title[]{Parameter estimation with non-stationary noise in gravitational waves data}
\correspondingauthor{Sumit Kumar}
\email{sumit.kumar@aei.mpg.de}
\author[0000-0002-6404-0517]{Sumit Kumar}
\AEI{}
\UniHannover{}
\author[0000-0002-1850-4587]{Alexander H. Nitz}
\AEI{}
\UniHannover{}
\author[0000-0002-8158-5009]{ Xisco Jim\'enez Forteza}
\AEI{}
\UniHannover{}


\begin{abstract}
The sensitivity of \ac{gw} detectors is characterized by their noise curves, which determine the detector's reach and ability to measure the parameters of astrophysical sources accurately. The detector noise is typically modeled as stationary and Gaussian for many practical purposes and is characterized by its \ac{psd}. However, due to environmental and instrumental factors, physical changes in the state of detectors may introduce non-stationarity into the noise. Misestimation of the noise behavior directly impacts the posterior width of the signal parameters. It becomes an issue for studies that depend on accurate localization volumes, such as i) probing cosmological parameters (e.g., Hubble constant) using cross-correlation methods with galaxies, ii) doing \ac{em} follow-up using localization information from \ac{pe} done from pre-merger data. We study the effects of dynamical noise on the \ac{pe} of the GW events. We develop a new method to correct dynamical noise by estimating a locally valid pseudo-\ac{psd} normalized along a potential signal's time-frequency track. We do simulations by injecting binary neutron star (BNS) merger signals in various scenarios where the detector goes through a period of non-stationarity with reference noise curves of third-generation detectors (Cosmic Explorer, Einstein telescope). As an example, for a source where mis-modeling of the noise biases the signal-to-noise estimate by even $10\%$, one would expect the estimated sky localization to be either under or over-reported by $\sim 20\%$; errors like this, especially in low-latency, could potentially cause follow-up campaigns to miss the actual source location. 
\end{abstract}

\section{Introduction}
Since the first detection of gravitational waves (GWs) from the merger of two compact objects in 2015 \citep{2016LVCGW150914}, GW science has come a long way, and GW detections have become a routine~\citep{LIGOScientific:2021djp, nitz20214ogc}. LIGO observatories in Livingston and Hanford~\citep{LIGOScientific:2014pky} have already finished three science runs, namely O1, O2, and O3~\citep{LIGOScientific:2021djp}. The GW observatories: Virgo~\citep{VIRGO:2014yos} in Italy and KAGRA in Japan~\citep{KAGRA:2020tym} joined the LIGO-Hanford and LIGO-Livingston observatories in O2 and O3 respectively. Currently, there are more than 90 confident GW detections from the merger of compact objects as identified by various data analysis pipelines summed up as GWTC-3 catalog \citep{LIGOScientific:2021djp} and 4-OGC catalog \citep{nitz20214ogc}.
\noindent
In the upcoming decade, the network of ground based GW detectors is set to be expanded with the addition of LIGO-India \citep{Saleem:2021iwi}. The current ground based detectors are also set to undergo phases of improvement in sensitivity \citep{KAGRA:2013rdx}. Additionally, the proposed third generation (3G) of ground-based detectors~\citep{Shoemaker:2021ura} such as the Einstein telescope (ET) \citep{Sathyaprakash:2012jk}, and Cosmic Explorer (CE) \citep{Reitze:2019iox, Evans:2021gyd} are expected to have an order of magnitude better sensitivity and will be able to probe lower frequencies ($f_\mathrm{low} \sim 3-5$ Hz).   The detectors' better overall sensitivity will help improve the signal's strength, while the better sensitivity at low frequencies will enable us to accumulate more signal from the inspiral phase of a binary merger. It will enable us to see farther into the universe; hence, we will have a rapidly expanding GW catalog in the 3G era.

The ever-growing GW catalog enables us to answer important scientific questions that are unique to GW observations, such as inferring the mass function, spin distribution, and redshift distribution of binary black hole (BBH) mergers  \citep{LIGOScientific:2020kqk}. By inferring the localization volumes of individual GW merger events with the help of \ac{pe} methods and by cross-correlating the localization volumes with galaxy catalogs, one can also estimate the cosmological parameters such as the Hubble constant \citep{LIGOScientific:2021aug}, and large scale distribution of matter \citep{Vijaykumar:2020pzn, Mukherjee:2020hyn, Libanore:2020fim, Mukherjee:2020mha, 2021Herrera}. The reach of third-generation detectors also implies a large population of loud BNS mergers, which will be prime candidates for \ac{em} follow-up campaigns. It will also be feasible to infer the sky location from the early part of the signal and alert the \ac{em} telescope before a merger and the subsequent \ac{em} emission occurs~\citep{Chan:2018csa, Nitz2021premerger}. The correct estimation of the localization volumes requires an accurate understanding and modeling of the noise of the detector along with the signals.
\noindent

The data from a network of GW detectors can be brought together and analyzed coherently as if they constituted a large synthetic detector, or it can be treated separately for each detector followed up by coincidence analysis \citep{Robinson:2008un}. Most search methods for compact binary coalescence (CBC) sources use coincidence methods \citep{Abbott:2020}, e.g., \texttt{PyCBC} \citep{Usman:2015kfa, Nitz:2017svb} and \texttt{GstLAL} \citep{Messick_2017}. The noise is typically assumed to be Gaussian, stationary, and colored. The noise of a GW detector can be characterized by its \ac{psd}, which quantifies the noise power present in each frequency bin. The general GW data analysis methods can be broadly categorized into two main groups: i) GW searches, where an entire time series of GW data are thoroughly examined for potential GW signals or triggers, and ii) \ac{pe}, which follows the searches where the top-ranked triggers are followed up and further analysis is carried out to estimate the parameters associated with the detected signals.

In GW searches, for each detector, the \ac{psd} of the noise is typically estimated for a segment of data $\sim256-1024$s using Welch's method by dividing it into overlapping shorter chunks~\citep{Abbott:2020, Usman:2015kfa}. It is assumed that the \ac{psd} is constant during this time. This \ac{psd} is then used to calculate the signal-to-noise ratio (SNR) time series $\rho(t)$ by applying matched filtering to the data for a bank of template waveforms~\citep{Allen:2005fk}. Whenever $\rho$ is obtained above a certain threshold, it is called a single-detector trigger. The local trigger density depends on the \ac{psd}. If a local deviation from the average \ac{psd} is not considered, it can affect the estimation of the local trigger density~\citep{Mozzon2020}.

For \ac{pe}, the \ac{psd} for each detector is estimated for a time window enclosing the GW signal. The data from each detector is used to calculate the multi-detector likelihood function. The \ac{psd} is assumed to be constant during the time of the signal in the detector frequency band. Any significant local time variation from the estimated \ac{psd} will affect the likelihood function and, therefore, the width of the posterior samples and, hence, the reported uncertainties in the marginalized distribution of individual parameters. For the current generation detectors, we do not expect this bias to be statistically significant for loud signals because most of the signals are short-lived ($\sim \mathcal{O}(1-10)$ seconds for BBH and $\sim \mathcal{O}(100)$ seconds for BNS) in the detector sensitivity band (20 Hz - 2000Hz). However, for third-generation detectors, we expect the signals to last for $~\sim \mathcal{O}(100-60000)$ seconds because of the better sensitivity at low frequencies. Due to very long signal durations, any change in noise characteristics due to evolution in the state of the detectors will cause the noise to be non-stationary; modeling the non-stationarity is needed to ensure the most accurate source localization and parameter estimates. Moreover, with forthcoming upgrades to the sensitivity of current ground-based detectors and their increasing capability to detect low-frequency signals, coupled with advancements in waveform models (including higher modes), the duration of signals within the detector frequency bands is expected to extend further. Therefore, incorporating non-stationary noise into \ac{pe} methods becomes crucial, emphasizing the need to develop appropriate techniques whenever such noise is present in the detector data.

An important aspect of long-duration signals is that a fraction of them can be identified from a few minutes to a few hours before the merger with sky localization information~\citep{Nitz2021premerger}. It can alert various telescopes to point in the right direction so they are ready to observe the \ac{em} follow-up observations. The very first detection of a BNS merger, GW170817, by the LIGO-Virgo detector network was observed simultaneously in gamma rays \citep{Goldstein:2017mmi, LIGOScientific:2017zic, Savchenko:2017ffs, LIGOScientific:2017zic}, and other observations in the EM spectrum such as in x-ray, optical, and radio frequencies \citep{LIGOScientific:2017ync}. With long-duration signals of interest such as BNS or neutron star-black hole (NSBH) mergers, we can design strategies to estimate the localization volumes sometime before the merger. A misestimation of localization volume for a pre-merger alert will end up in either wasting telescope time and resources (in case of over-estimated localization volume) or, in the worst case scenario, missing out on observing the event (underestimated localization volume).

There are studies focused on challenges in developing GW \ac{pe} methods in the presence of non-stationary noise, such as \cite{Edy2021}. Traditional approaches to incorporate non-stationarity make the computation of the likelihood function very expensive. This paper explores the modeling of non-stationary noise and its implications on \ac{pe}. We introduce an efficient method to account for non-stationarity, which slowly varies compared to signal evolution. It is accomplished by calculating a corrected \ac{psd} along the track of the signal by dividing the time-frequency track of a signal into shorter duration segments and estimating a local \ac{psd} for each time bin separately. The corrected \ac{psd} can be constructed by selecting the frequency range from each local \ac{psd} estimate corresponding to the expected signal contribution in the same time range. This method of estimating the \ac{psd} depends on obtaining a preliminary constraint from searches for trigger parameters to identify the signal's track in time-frequency space.

This paper is organized as follows: Section II discusses the noise properties of GW detectors. We discuss the Gaussian and stationary noise assumption in the absence of transient noise. We then define the non-stationarity of the noise in terms of the statistical moments. The methods to test the assumptions of Gaussianity and stationarity are also highlighted. We also discuss the effects of misestimation of \ac{psd} on inferred signal properties. In section III, we discuss the \ac{psd} variation statistic, which can be used to estimate the local variation in the noise of the detector and quantify the non-stationarity. We show examples of induced non-stationarity in simulated data and methods to recover them using the \ac{psd} variation statistic. Section IV presents a method to estimate a `corrected pseudo-\ac{psd}' and apply it to some realistic injected signals for third-generation detectors such as the Einstein telescope. We show how a corrected \ac{psd} can help us recover the corrected SNR. We then simulate multiple BNS injections in a network of third-generation detectors consisting of an Einstein Telescope and two Cosmic Explorer with non-stationary noise. We do the \ac{pe} with corrected (and uncorrected) \ac{psd} and show our results. We summarize our results and conclusions in section V.

\section{GW data analysis and Noise Characterization}
The signal from the merger of two compact objects, as seen in a detector, consists of a time series $h(t)$ buried in the noisy data of the detector $s(t)$.

\begin{equation}
    s(t) = 
 \begin{cases}
    n(t) + h(t),& \text{When signal is present}\\
    n(t),              & \text{otherwise}
 \end{cases}
\end{equation}
\noindent
The shape of the signal $h(t)$ depends on the intrinsic parameters of the source, such as the component masses ($m_{1,2}$), the component spins ($s_{1x,2x}, s_{1y,2y},s_{1z,2z}$) as well as on the extrinsic parameters such as the luminosity distance to the source ($D_L$), the inclination angle of the binary to line of sight ($\iota$), the polarization angle ($\psi$), the coalescence phase ($\phi$), the sky localization (right ascension: $RA$, declination: $dec$). The noise $n(t)$ has contributions from many sources, such as quantum noise, thermal noise, Newtonian noise, and transient noise of unknown origins \citep{Abbott:2020}. The overall noise, in the absence of transient noise, is typically described as a stochastic process modeled as a Gaussian distribution:
\begin{equation}
    p(\mathbf{n}) = \frac{1}{\sqrt{2\pi |C_{ij}|}}\exp(-\frac{1}{2}\sum_{i,j}(n_i - \mu) C_{ij}^{-1} (n_j - \mu))
    \label{eq:GaussianNoiseModel}
\end{equation}
Where $\mathbf{n}$ is a time series vector of a certain noise realization with a given sample rate. $\mu$ is the expectation value of the noise and $C_{ij}$ is the covariance between $i^{th}$ and $j^{th}$ bin in the time series. Both these quantities can be estimated from the data. The noise is said to be stationary if the covariance matrix $C_{ij}$ only depends on the lag between bins $i$ and $j$ i.e. $|t_i - t_j|$. A useful property of stationary noise is that the covariance matrix takes a positive diagonal form in the frequency domain, i.e., $C_{ij} = \delta_{ij} S_n(f_i)$. Here $S_n(f_i)$ is defined as \ac{psd}, which is the Fourier transform of the correlation function $C(\tau)$ where $\tau = |t_i-t_j|$ is the time lag. Non-stationary noise is defined as the noise in which statistical moments such as expectation value of noise $\mu$, and (auto) correlation $C_{ij}$, are not stationary. In other words, they become an explicit function of time $t$.

The validity of the assumptions of the noise being Gaussian and stationary in the detector can be tested by various methods, e.g., by measuring the noise covariance of the data \citep{Edy2021}. One can also use the expected local variation in \ac{psd} characterized by Gaussian and stationary noise to look for chunks in the data inconsistent with these assumptions \citep{Mozzon2020}.  Since a signal is buried in a particular noise realization of the detector and we can only know the statistical properties of the noise, we always recover a projected signal in signal+noise space. The recovery of the signal has an offset compared to the actual signal parameters. This offset depends on the particular noise realization and shall be consistent with the distribution of the expected offset around the true value. This distribution of expected offsets can be found analytically or through simulations for the ideal case of Gaussian and stationary noise. In \cite{kulkarni2021reliability}, authors present another method to test the assumptions of Gaussian and stationary noise in the presence of the signal: one can launch an injection-recovery campaign in the vicinity of the signal assuming the noise properties do not change in the chosen window. By studying the recovered parameters from many injections, it can be determined if the recovered credible intervals match the statistical expectations of Gaussian and stationary noise \citep{kulkarni2021reliability}.

In the scenario where detector data have non-stationary noise, but we still use the assumption of Gaussian and stationary noise, the estimated \ac{psd} with the traditional Welch method may not be a correct representation of noise properties. In other words, the \ac{psd} of the detector may be misestimated.

\subsection{Impact of misestimating the \ac{psd} on signal property Inference}
 In this section, we will discuss the effects of the misestimation of \ac{psd} on the estimated signal properties. We focus on the scenario where \ac{psd} is not evaluated correctly. It can happen either due to i) not accounting for the non-stationary nature of noise, if such non-stationarity is present, or ii) local statistical fluctuation in the noise when a much longer segment is used for \ac{psd} estimation, compared to the signal duration. We now review some basic concepts in GW searches and \ac{pe}. The signal-to-noise ratio (SNR) of a signal template $h$ is defined as $\rho \equiv \frac{(s|h)}{\sqrt{(h|h)}}$ where the inner product $(a|b)$ between the two time series $a(t)$ and $b(t)$ (and their counterpart in Fourier domain $a(f)$ and $b(f)$) is defined as:
\begin{equation}
    (a|b) \equiv 2\int_{-\infty}^{\infty} \frac{a(f)b^{\star}(f)}{S_n(|f|)}, \label{eqn:inner_product}
\end{equation}
\noindent
where $S_n(f)$ is the estimated power spectrum of the noise over a chosen time window. In GW search pipelines, an SNR time series is constructed for given template waveforms in each detector. For each pair of detectors in the network, single-detector triggers are subjected to a `coincidence test' where only those pair of triggers, which are within the light travel time (plus a few milliseconds to account for errors in estimating trigger time) between the two detectors, survive. The end result of a search pipeline is to statistically rank the GW triggers with inverse false alarm rate from most significant event to least compared to the noise background \citep{Usman:2015kfa}.

As the estimation of SNR depends on estimated \ac{psd} $S_n(f)$, misestimation of \ac{psd} will affect the SNR and, hence, the trigger density. There are methods available to correct for these effects at the level of the searches, such as applying a \ac{psd} drift correction \citep{Zackay2021} and applying dynamic normalization using the \ac{psd} variation statistic \citep{Mozzon2020}. However, note that both methods assume a constant change in the overall amplitude of the noise and that this is constant throughout the duration of the signal, i.e., a general variation of amplitude as a function of frequency is not accounted for. We may expect this to be a reasonable assumption for short-duration signals ($\sim\mathcal{O}(1-10)$ seconds) such as BBHs.

For the \ac{pe}, we describe the signal $h(\bm{\theta})$ in terms of parameters $\bm{\theta}$ of the waveform model and perform Bayesian analysis to obtain the posterior distribution $p(\bm{\theta} | d)$ which represents the probability distribution over the parameters $\bm{\theta}$ given the observed data $d$ and model,

\begin{equation}
    p(\bm{\theta} | d) = \frac{\mathcal{L}(d | \bm{\theta}) \pi(\bm{\theta})}{p(d)}, \label{eqn:bayes}
\end{equation}
\noindent
where $\mathcal{L}(d | \bm{\theta})$ is the likelihood function: the probability of obtaining the data given the parameters $\bm{\theta}$ describing the model. $\pi(\bm{\theta})$ is the prior probability distribution for the model parameters $\bm{\theta}$, and $p(d)$ is known as ‘Bayesian evidence’ or ‘marginalized likelihood’ which acts as a normalization factor. For the Gaussian noise model, the likelihood function in the frequency domain takes the form:

\begin{equation}
\mathcal{L} \propto \exp{-\frac{1}{2}(\mathbf{s}-h(\bm{\theta})|\mathbf{s}-h(\bm{\theta}))}, \label{eqn:likelihood}
\end{equation}
where $\mathbf{s}$ is the strain data and $h(\bm{\theta})$ is the template WF model. The inner product $(.|.)$ is defined in equation \ref{eqn:inner_product}. 
As shown in \citep{Edy2021}, the effects of misestimating the \ac{psd} can be calculated by doing a linear series expansion of the waveform template $h$ around the true signal value $h_{0}$. The misestimation of the $S_n(f)$ only results in over (or under) estimating the posterior width. In other words, if we use two different estimates of $S_n(f)$, the posteriors will still be centered at the same values, but the width of the posteriors will be different. Specifically, at linear order in the signal $h_{0}$, the covariance matrix of the waveform parameters -- the inverse of the Fisher matrix~\citep{Vallisneri:2007ev} -- scales linearly with $S_n(f)$, while the SNR of the signal as $S_n^{-1/2}(f)$. Thus, as a rule of thumb, a \ac{psd} variation of type $S_n(f)\rightarrow \alpha S_n(f)$ will broaden/narrow the width on the posterior distributions of the physical parameters by the factor $\sqrt{\alpha}$, which results in changing their inferred uncertainty. On the other hand, the non-stationary models considered are not expected to introduce any bias on the estimated parameters, at least at linear order~\citep{Edy2021}.

The accurate width of posterior samples may become relevant for studies where the knowledge of correct localization volumes is critical, such as i) pre-merger localization, ii) estimating Hubble parameter using cross-correlation methods with galaxy catalogs, and iii) probing large-scale structures using two-point correlation function of localization volumes with next generation GW detectors. For all the studies mentioned above, estimating the correct localization volume is essential for unbiased estimators. For the Hubble constant example, a study by \cite{mozzon2021does}, through simulation of BNS signals, found that though non-stationary noise will not be a limiting factor for resolving Hubble tension for the current generation of detectors, it will play an important role for 3G detectors.

If the data are non-stationary, the covariance matrix in the frequency domain also contains the off-diagonal elements, and the estimation of the likelihood function, equation \ref{eqn:likelihood}, becomes computationally expensive. So, it is difficult to model the non-stationarities straightforwardly without compromising the computational costs. In one of the studies, \citep{Edy2021} shows that it is crucial to account for the non-stationarity in the signals that last longer than a few minutes, such as BNS signals in current detectors and BBH/BNS signals in third-generation GW detectors.

The methods used in estimating \ac{psd} can have statistical uncertainties associated with it. It is essential to account for these uncertainties and their impact on \ac{pe} \citep{2020PhRvR...2d3298T,2020PhRvD.102b3008B, 2021PhRvR...3d3049T, 2019PhRvD.100j4004C}. Calibration uncertainties also affect the likelihood function that needs to be included while dealing with the actual data \citep{SplineCalMarg-T1400682, Cahillane:2017vkb}. In this work, we only focus on the non-stationarities in the estimation of the \ac{psd} and on subsequent \ac{pe}.

\section{Estimating local variation in the PSD}
\noindent
In order to test the Gaussianity and stationarity of the data or to look for the non-stationary chunk of the data, one can employ one of the many methods described above, such as estimating the noise covariance, or use the \ac{psd} variation statistic, $v_s$ described in  \citep{Mozzon2020}. Here, we briefly discuss the \ac{psd} variation statistic.
The non-stationarity of the data will lead to a difference in actual power spectrum $S_A(f)$ and estimated one $S_E(f)$ modeled with the help of the \ac{psd} variation statistic $v_s$,
\begin{equation}
    S_A(f) = v_s S_E(f)
\end{equation}
\noindent 
Using a filter $\mathcal{F}(f) = \frac{|h(f)|}{S_E(f)}$, with approximating $|h(f)| \propto f^{-7/6}$ (dominant amplitude behavior of the CBC signal during the inspiral phase), $v_s$ can be estimated by computing the convolution between the data $s(f)$ and filter $\mathcal{F}(f)$.

The non-stationarity in the detector noise can be attributed to various factors of known or unknown origin. Different non-stationary noise models are considered in the literature \citep{PhysRevD.101.042003, hebbal2019bayesian, Edy2021}. A few examples are:
\begin{itemize}
    \item \textit{local fluctuations in the detector noise due to non-stationary component:} It might be the case that there is some non-stationary component in the overall noise. This can be modelled as follows: $n(t) = An_1(t) + B(t) n_2(t)$ \citep{Edy2021}. Here, the first component $n_1(t)$ is stationary Gaussian noise with normalization constant A, and the second component $n_2(t)$ represents noise of unknown origin with a time-dependent component $B(t)$. A special case of the above is when the noise is Gaussian, but the overall amplitude varies as a function of time such that $n(t) = (1+B(t))A_0n_1(t)$, where $A_0$ is a constant.
    \item \textit{A shift in the overall \ac{psd}:} At times, the detector may go offline (out of science mode) and then return to operation, but the overall \ac{psd} does not stabilize. Even during uninterrupted observation runs, fluctuations in the detector's \ac{psd} can be observed. The overall variation in the \ac{psd} can be modeled as a constant multiplied by the reference \ac{psd}. Over time, it gradually returns to a more stable configuration resembling the original reference \ac{psd}. This type of local fluctuations of $\sim\mathcal{O}(10)$ seconds can be found in the non-stationary chunks considered in the study of \citep{Mozzon2020}. A specific instance of this is when the general `noise floor' of the \ac{psd} is elevated and then settles back to the original \ac{psd}. Mathematically, this situation represents a special case of non-stationary noise described in the previous section, with a specific form of the function $B(t)$.
\end{itemize}

\begin{figure*}
    \centering

    \begin{minipage}{0.48\textwidth}
        \centering
        \includegraphics[width=\linewidth]{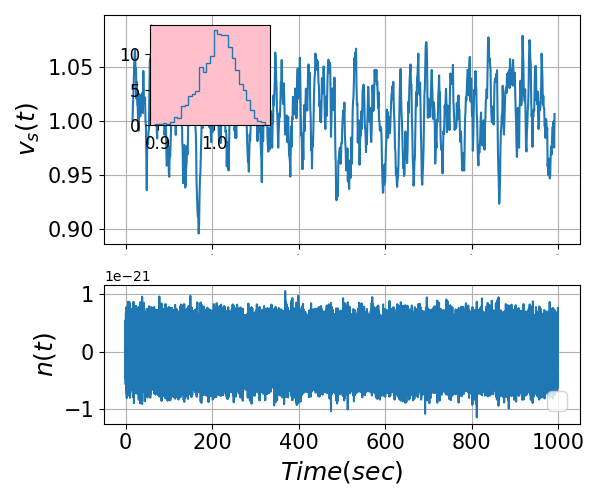}
    \end{minipage}
    \hfill
    \begin{minipage}{0.48\textwidth}
        \centering
        \includegraphics[width=\linewidth]{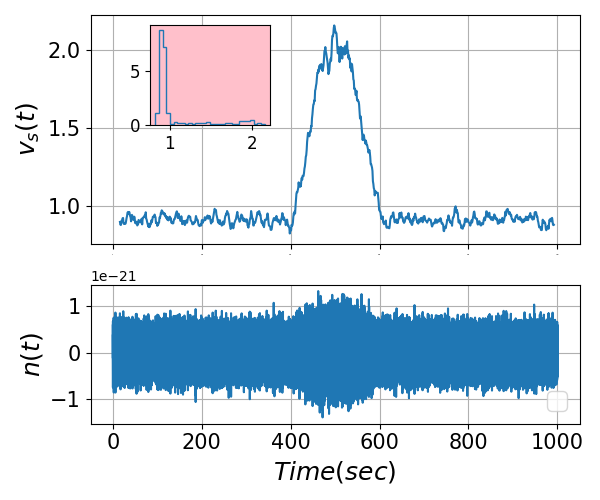}
    \end{minipage}

    \vspace{1em}

    \begin{minipage}{0.48\textwidth}
        \centering
        \includegraphics[width=\linewidth]{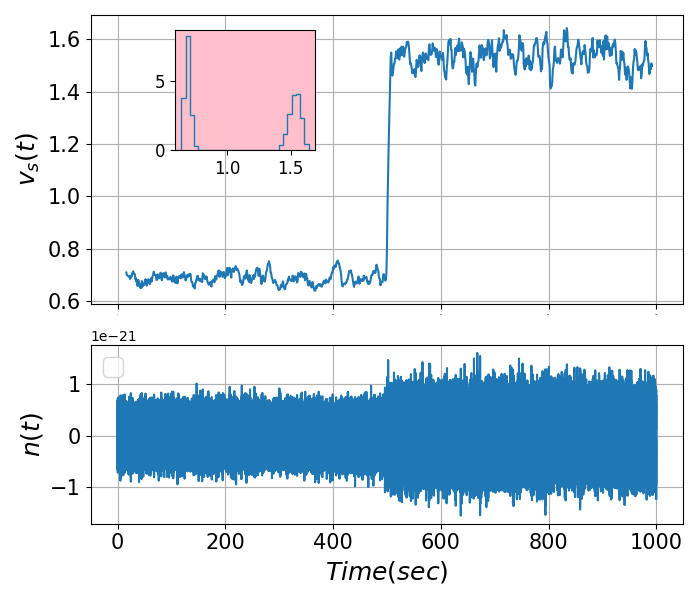}
    \end{minipage}
    \hfill
    \begin{minipage}{0.48\textwidth}
        \centering
        \includegraphics[width=\linewidth]
        {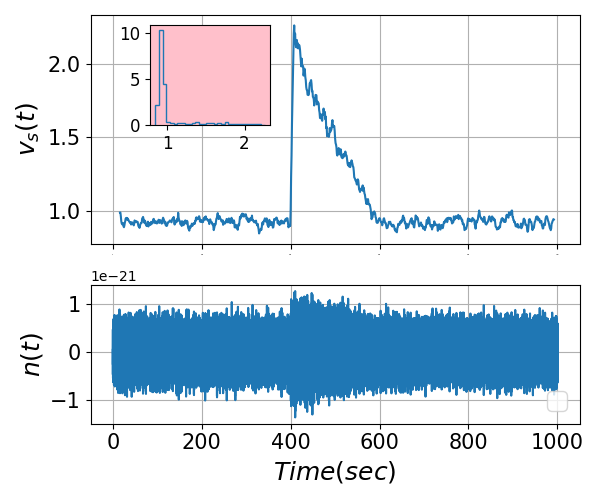}
    \end{minipage}
    \caption{Different types of local variation in the \ac{psd} are considered here. We use the \ac{psd} variation statistic $v_s$ described in \citep{Mozzon2020} to track the local variation in \ac{psd}. The lower panel shows the strain data for 1000 seconds in all the figures. The top panel shows the local \ac{psd} variation statistic with respect to the \ac{psd} estimated for the entire duration (1000 seconds). The inset shows the distribution of all the \ac{psd} variation values. Except for the top left panel, these figures consider different scenarios of non-stationarities: (a) Top left: Baseline noise model with Gaussian and stationary assumptions. The local variations in the \ac{psd} are due to the Gaussian random fluctuations. (b) Top right: A filter with a bump in the middle acts upon the baseline noise model. We use $B(t) = B_0 sin(\frac{\pi}{\Delta T} t)$ with $[B_0,\Delta T]$ = $[0.5,200 sec]$ so that it produces a bump in the amplitude of positive half sine wave between t=400 to t=600 seconds. (c) Bottom left: The floor of the baseline \ac{psd} is lifted in the middle (at t=500 seconds) with $B(t)=0.5 ~\text{for}~ t>500$sec. (d) Bottom right: The floor of the \ac{psd} is lifted at t=400 seconds and drops down to the base scale slowly in due time by t=600 seconds.}
    \label{fig:psd_examples}
\end{figure*}

\subsection{Understanding non-stationary noise with toy models}
We simulate a few examples of non-stationary noise scenarios without the presence of the signal. In figure \ref{fig:psd_examples}, using toy models, we show some examples of the abovementioned non-stationarities. We use a baseline noise model (Gaussian, stationary colored noise) and apply a filter to scale the amplitude of Gaussian noise as a function of time to generate noise realization with the desired non-stationary nature. It may not be the most general way to model non-stationarity, but we still consider it as a test study. For the examples shown in the figure, we generated $\sim 1000$ seconds of data with Gaussian noise with \texttt{aLIGOZeroDetHighPower}, one of the available noise models in the publicly available software package \textsc{LALSuite}. We call this the baseline noise model. The top left panel in the Figure \ref{fig:psd_examples} represents a realization of the baseline model. We use the \ac{psd} variation statistic $v_s$ to track the non-stationary nature of the noise. For the baseline model, the local \ac{psd} variation is due to purely Gaussian random fluctuations, so the \ac{psd} variation statistic $v_s$ has a mean around $<v_s> = 1$ with a standard deviation consistent with the Gaussian fluctuations. We generate the non-stationary noise by applying a filter on baseline noise such that the amplitude of the Gaussian noise varies as desired. This type of \ac{psd} variation can be modeled as
\begin{equation}
    n(t) = (1 + B(t)) A_0 n_G(t), \label{eqn:noise_model}
\end{equation}
where $A_0n_G(t)$ is the stationary, Gaussian, colored noise, $B(t)$ models the variation in the amplitude. $B(t) = 0$ corresponds to the baseline noise model. As shown in figure \ref{fig:psd_examples}, we introduce some arbitrary non-stationarity in the noise at various times and use \ac{psd} variation statistic $v_s$ to track it. Since we know the expected distribution of $v_s$ for Gaussian and stationary noise, we can use $v_s$ to track the chunks in the data that are inconsistent with the baseline model. In figure \ref{fig:psd_examples}, we show three toy models of non-stationary noise: 
\begin{itemize}
    \item A non-stationary bump in the strain noise between 400-600 seconds. To represent this, we introduce $B(t)$ as a half-sine cycle within 400-600 seconds. During this window, the strain data exhibits non-stationarity.
    \item The overall amplitude of the noise is multiplied by a constant function at t=500 seconds. In this case, the first and second half of strain data can be considered stationary on their own. However, since the noise variance is not the same for different segments, the overall data can be considered non-stationary.
    \item The noise amplitude is multiplied by a constant factor at t=400 seconds and then gradually goes down to the original noise at t=600 seconds. It can be modeled by a linear function  B(t). 
\end{itemize}

In the examples presented in Figure \ref{fig:psd_examples}, we have demonstrated how the \ac{psd} variation statistic, denoted as $v_s$, can be utilized to verify whether the noise characteristics of the GW detector data align with the assumptions of Gaussianity and stationarity. In these cases, the \ac{psd} variation statistic $v_s$ correlates with the amplitude $B(t)$ described in Equation \ref{eqn:noise_model}. It allows us to track the specific nature of the non-stationarity in these instances. The following section will explore how non-stationarity during a GW signal impacts the single-detector SNR and \ac{pe} when employing multiple detectors.

\section{Modeling the non-stationarity for parameter estimation}
In most of the standard \ac{pe} pipelines, to obtain the posterior distribution, equation \ref{eqn:bayes}, a sampling method (such as Markov Chain Monte Carlo, nested sampling) is used with a likelihood model and prior probability distribution over the model parameters. The \ac{psd} is estimated for a segment of the data around the signal using Welch's method with smaller overlapping segments or spline-modeling such as implemented by the Bayesline algorithm~\citep{Littenberg:2014oda}. For example, to estimate the \ac{psd} for \ac{pe} in the 4OGC and 3OGC analyses \citep{nitz20214ogc, nitz20213ogc}, 512 seconds of data centered around the trigger are used to estimate the \ac{psd} with overlapping segments of 8 seconds (unless the data are corrupted in this window, in which case, this window is shifted and/or shrunk). The estimated \ac{psd} is then used to calculate the likelihood function, equation \ref{eqn:likelihood}. The \ac{psd} averaged over the long segment surrounding the signal does not capture local fluctuations (or non-stationarities) on a time scale much shorter than the averaging window that might be present along the signal's track, as the averaging process flattens the local fluctuations in the \ac{psd}. To account for non-stationarity in the \ac{pe} framework, the standard method outlined in \citep{Edy2021} requires a redefinition of the likelihood function. It involves incorporating off-diagonal terms into the covariance matrix estimated in the frequency domain. However, this approach makes the computation of the likelihood function significantly more expensive.

In this work, we propose a new method to model specific types of non-stationarities where the strain data can be divided into smaller chunks of `stationary' or `semi-stationary' parts, such as slowly varying Gaussian noise amplitude, sudden lift or drop in the noise floor, even missing some duration of data (for example detector being offline) for very long duration signals. For 3G GW detectors such as ET and CE, due to their low-frequency reach, the BBH signals can be $\sim \mathcal{O}(100)-\mathcal{O}(1000)$ seconds long and BNS signals can be  $\sim \mathcal{O}(1000)-\mathcal{O}(60000)$ seconds long depending on the value of low-frequency cutoff ($f_\mathrm{low} \in [2,5]$ Hz). During this time, the detector may go through periods of non-stationarities. Even for the current generation detectors, a BNS signal in the detector band is usually a few hundred seconds long. When the duration of the signal within the detector band is sufficiently long, estimating the \ac{psd} along the signal's track becomes feasible. We can utilize the reference signal parameters obtained from the GW search pipelines to achieve this. Specifically, we can select the template with the highest signal-to-noise ratio (SNR) as our reference signal. We describe the important steps of the algorithm as follows:
\begin{figure}[t]
\includegraphics[width=0.49\textwidth]{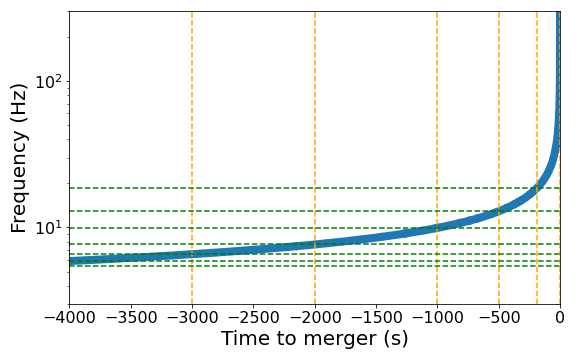}
\caption{We show a time-frequency track of a GW signal from the merger of BNS. We use the track of the reference signal to identify the times corresponding to the frequency bins to estimate the \ac{psd} of the noise. We specify corresponding frequency ranges for the timestamps of the signal lying in the non-stationary noise, and a binning scheme can be determined accordingly.}
\label{fig:signal_track}
\end{figure}

\begin{itemize}
\item Create the track of the signal in the time-frequency plane. In figure \ref{fig:signal_track}, we show a reference BNS signal's time-frequency track.
\item Utilize the \ac{psd} variation statistic $ v_s $ to identify the timestamps in the detector corresponding to periods of non-stationary noise. To achieve this, we compare the $ v_s $ time series to the expected distribution of $ v_s $ values. For the examples discussed in this work, we use a cutoff of $ v_s = 1.2 $ to flag non-stationary periods above this threshold.
\item Based on the time-frequency evolution of the reference signal, we identify the frequency range of the signal that corresponds to the non-stationary data. During the period of non-stationarity, the size of each time bin should be sufficiently large to allow for an accurate estimation of the \ac{psd} within that bin.
\item Estimate the \ac{psd} for each frequency bin using strain data from the corresponding time bin and stack the estimated \ac{psd} in each frequency bin to estimate a corrected \ac{psd} along the signal track.
\end{itemize}

\begin{table*}
    \footnotesize
    \caption{The description of various scenarios of non-stationary noise is considered. We use a noise sensitivity curve for Cosmic Explorer design sensitivity \citep{Hall:2020dps} with a low-frequency cutoff of $f_\mathrm{low} = 5Hz$. The non-stationarity in the noise is introduced by applying a filter to the baseline noise model with a time-varying amplitude for a fixed time window. To estimate the SNR, we estimate the \ac{psd} by two methods: i) Using a long segment of 8000 seconds surrounding the trigger time $t_c$, using Welch's method by dividing the whole segment into smaller chunks of 8 seconds. And ii) estimated along the time-frequency track of the reference BNS signal. The \ac{psd} estimated along the track of the signal is used as a reference to estimate the fractional difference in SNR $=\abs{\delta \rho} / \rho_{\text{ref}}$. All the examples shown here are single detector examples. We quote two SNR differences: In one, the reference SNR is estimated with a corrected \ac{psd} using our method. We also provide the expected difference, where the reference SNR is estimated using a \ac{psd} calculated analytically from the known injected variation. The difference between our method and the expected values may stem from two main factors: i) random fluctuations in the noise realization and ii) large error bars in estimating the \ac{psd} in smaller time bins. This discrepancy is also illustrated in the \ac{psd} estimation compared to the expected \ac{psd} in Figure \ref{fig:psd_estimation_examples}.
    }
    \label{table:nonstationary_scenarios}
    \begin{center}
    \noindent\setlength\tabcolsep{3pt}%
    \begin{tabularx}{\textwidth}{llp{0.3\textwidth}cc}
    \toprule \hline
    \multicolumn{1}{c}{\multirow{2}{*}{S. No.}} & \multirow{2}{*}{Amplitude variation} & \multirow{2}{*}{Description} & \multicolumn{2}{c}{SNR difference}  \\
\multicolumn{1}{c}{}                   &                    &                    & Our Method & Expectations \\
    \hline
    1. &\centering \( B(t) = \begin{cases}
        B_0,& : t > t_\mathrm{shift} \\
        0,              & :\text{otherwise}\end{cases}
    \) & A positive shift, represented by a step function, occurs in the amplitude of the Gaussian noise at $t_\mathrm{shift} = t_c - 190 $ seconds. Here, $B_0 = 0.8$, and $ t_c $ denotes the trigger time of the signal. Therefore, for the 190 seconds leading up to the merger, the signal remains buried in the noise, characterized by a \ac{psd} where the noise floor is elevated. &  43.6 \% & 35.3 \% \\

    2. &\centering \( B(t) = \begin{cases}
        -\frac{B_0}{\Delta T} (t-t_1) + B_0 , & :t_1  < t < t_1+\Delta T \\
        0,              & :\text{otherwise}\end{cases}
    \) & A positive shift in the amplitude of the noise followed by a linear decay to original amplitude (base scale). We use $B_0 = 0.8$, $\Delta T = 600$ seconds, and $t_1 = t_c - 650$. The signal spent 600 seconds of time in non-stationary noise where the noise floor was above the base scale. & 11.47 \% & 5.38 \%\\

    3. &\centering \( B(t) = \begin{cases}
        B_0\sin{(\frac{\pi (t-t_1)}{\Delta T})} ,& :t_1 < t < t_1+\Delta T \\
        0,              & :\text{otherwise}\end{cases}
    \) & A bump described by a half positive cycle of sine function with $B_0 = 0.8$,  $\Delta T = 600$ seconds, and $t_1 = t_c - 650$. The signal spend 600 seconds of time in non-stationary noise where noise floor was lifted above the base scale.  & 21.19 \%  & 11.07 \% \\

    4. &\centering \( B(t) = \begin{cases}
        B_0, & : t < t_\mathrm{shift} \\
        0,              & :\text{otherwise}\end{cases}
    \) & A drop in the amplitude of the Gaussian noise at $t_\mathrm{shift} = t_c - 10$ seconds. $B_0 = 0.8$, and $t_c$ is the trigger time of the signal. So the signal spend all the time until 10 seconds prior to the merger in the noise characterized by where noise floor is raised with respect to base scale. & 9.11 \% & 15.49 \%\\
    & & & & \\
    \hline
    \bottomrule
    \end{tabularx}
    \end{center}
\end{table*}
We must consider several specific factors when selecting the time-frequency binning for our analysis. We assume that the variations in the detector noise within each time (or frequency) bin are negligible, allowing us to model the noise in each bin as Gaussian and stationary. In other words, our binning scheme should ensure that even in non-stationary chunks, the \ac{psd} fluctuations within each smaller bin remain negligible. The bin size must also be sufficiently large to accurately estimate the \ac{psd} for each frequency bin, which limits the smallest bin size to a few seconds. Consequently, this method is unable to resolve non-stationarities that last shorter than a few seconds. Additionally, if the variations in noise properties occur on a faster timescale within each bin, we cannot assume that the data within that bin are semi-stationary or stationary. While this algorithm is well-suited for slowly varying non-stationary data, particularly for third-generation (3G) detectors, the current generation of detectors has encountered non-stationary intervals lasting approximately $\sim\mathcal{O}(10)$ seconds. For low-mass systems, such as binary neutron star (BNS) and neutron star-black hole (NSBH) mergers, the events can last a few hundred seconds in the detector frequency bins. If a non-stationary period of several tens of seconds overlaps with these signals, we can utilize this method to construct an effective \ac{psd} along the signal's trajectory. 

We simulate some examples of BNS signals injected in the detector data, which go through a period of non-stationarity. In Table \ref{table:nonstationary_scenarios}, we list the details of these simulations with arbitrary non-stationary noise. We inject a BNS signal with parameters ($m_\mathrm{1, source}$ = $1.5 M_\odot$, $m_\mathrm{2, source}$ = $1.3M_\odot$, $D_L$ = $100$ Mpc) in the data such that the detector goes through a period of non-stationarity during the signal. We estimate the SNR using two estimated \ac{psd}s: i) assuming the detector noise is constant during the signal and estimating average \ac{psd} from a long stretch of data around the signal, and ii) estimating \ac{psd} along the track of the signal as described above. We estimate the fractional SNR difference $( \frac{\delta_\rho}{\rho} )$ arising from two \ac{psd}s and found that there can be significant fractional SNR difference (up to $\sim 40\%$) for the arbitrary non-stationary noise we consider here. These differences might not be as extreme in real-life scenarios, but even a small difference in fractional SNR will affect localization volumes. We use the baseline noise model from the design sensitivity of the 3G detector cosmic explorer. We use the 3G detector noise curve only because BNS signals are expected to last for a very long time in the detector frequency band, and therefore, it serves as a good example. This method can also be applied to current detectors whenever we identify a non-stationary chunk of data (similar to the ones considered in the examples) along the track of a long BNS signal lasting a few hundred seconds. 

\begin{figure*}
    \centering
    \subfigure[]{\includegraphics[width=0.49\textwidth]{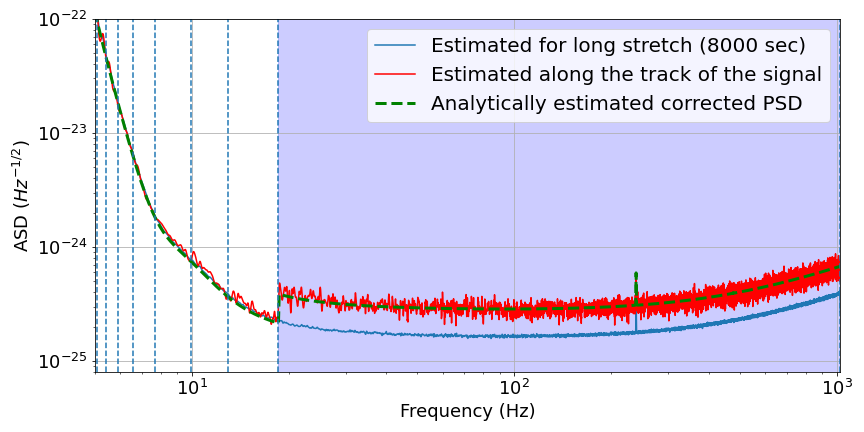}} 
    \subfigure[]{\includegraphics[width=0.49\textwidth]{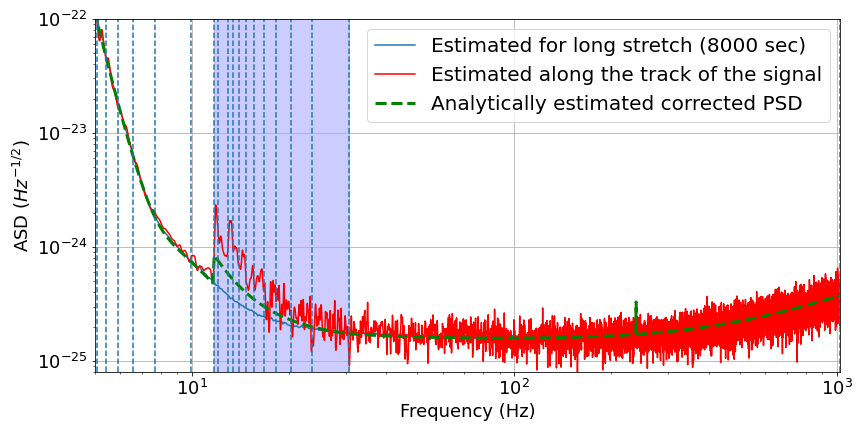}} 
    \subfigure[]{\includegraphics[width=0.49\textwidth]{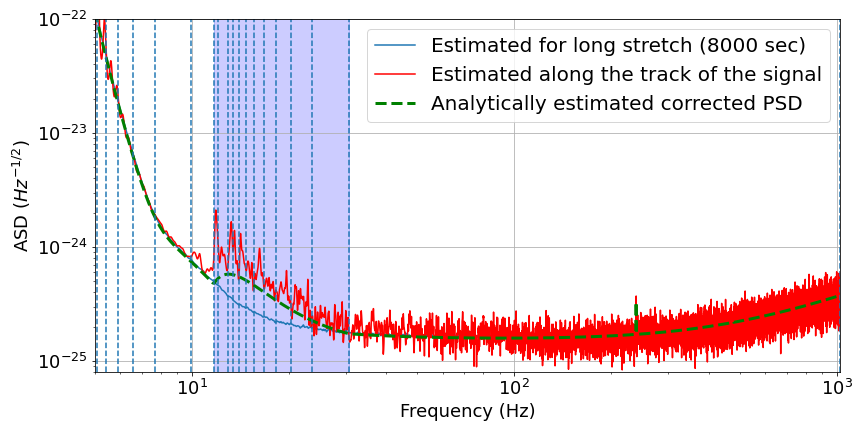}}
    \subfigure[]{\includegraphics[width=0.49\textwidth]{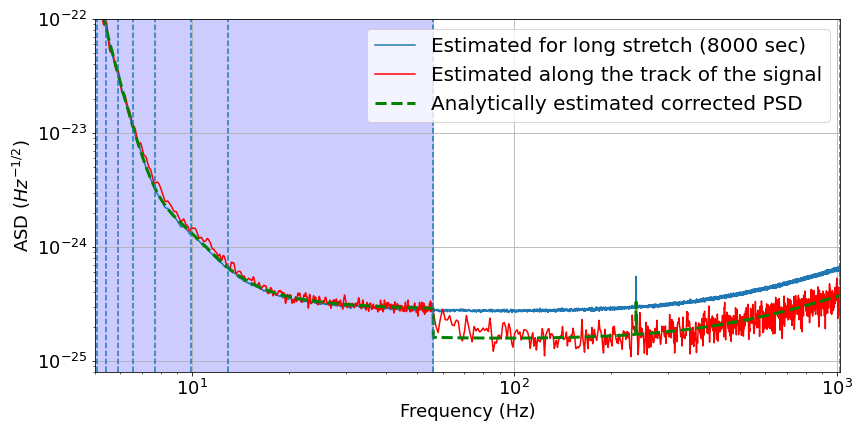}}
    \caption{The estimated amplitude spectral density (ASD) is shown as a function of frequency. The blue curve represents the ASD estimated for a long segment ($~8000$ seconds) around the signal, and the red curve represents the ASD estimated along the signal track as described in the text. We inject the BNS signal in the non-stationary noise as described in Table \ref{table:nonstationary_scenarios}. The shaded region represents the frequency range along the signal track when the detector went through the period of non-stationary noise. During this time, the noise amplitude increases as a time function. This is reflected in the raised (estimated) noise floor. Figures (a),(b),(c), and (d) correspond to scenarios (1),(2),(3), and (4) in the table \ref{table:nonstationary_scenarios} respectively. The difference in the estimated SNR for examples in table \ref{table:nonstationary_scenarios} arises due to the difference in the \ac{psd}s. In order to estimate the `correct' SNR in a non-stationary noise case, we need to use the \ac{psd} estimated along the signal track. The green dashed line represents the ASD along the signal track in each scenario. This analytical ASD is calculated based on our understanding of the non-stationarities introduced in the detector. The proximity of the green dashed line to the red curve indicates the effectiveness of our method in estimating the effective ASD along the signal's track.}
    \label{fig:psd_estimation_examples}
\end{figure*}

In Figure \ref{fig:psd_estimation_examples}, we present the estimated Amplitude Spectral Density (ASD), which is the square root of the \ac{psd}, for all the injection scenarios outlined in Table \ref{table:nonstationary_scenarios}. The frequency and time bins have been selected to capture the local variations in the \ac{psd}. We also include an expected ASD along the signal track to illustrate the algorithm's effectiveness. The expected ASD along the signal's trajectory is estimated analytically as $ \sqrt{S_\mathrm{track}(f)} = B(t(f))\sqrt{S_0(f)}$, where $B(t(f))$ is the amplitude variation described in table \ref{table:nonstationary_scenarios} and $t(f)$ is taken from the time-frequency track of the signal representing a time $t$ when the signal template hits the corresponding frequency $f$.

Recent improvements in \ac{pe} techniques, such as the implementation of heterodyned likelihood or relative binning methods \citep{Cornish:2010kf, 2021PhRvD.104j4054C, Zackay:2018qdy, Finstad:2020sok}, make fast \ac{pe} possible. It will make it viable to have pre-merger alerts in some situations where we have long signals. There also exist other methods, such as an alternative time-frequency GW data analysis approach that uses discrete, orthogonal wavelet wave packets \citep{2020PhRvD.102l4038C}, which can be used to quantify the non-stationarities and the \ac{psd} estimated (in case of non-stationary noise) can be used with the method presented here for \ac{pe}.

\subsection{Simulations and Parameter Estimation}
Equipped with all the tools described above, we can estimate the \ac{psd} along the track of the signal and use it in the \ac{pe}. When a signal stays in the frequency band of detectors, non-stationarity might be present in only one or multiple detectors at different times. To put this method to the test, we inject a GW signal from the merger of a BNS ($m_\mathrm{1, source}$ = $1.5 M_\odot$, $m_\mathrm{2, source}$ = $1.3 M_\odot$, $D_L$ = $1000$ Mpc) in a network of third generation gravitational wave detector. We consider two CE detectors (one located in the USA and the other in Australia) and one ET (Europe). We summarize the properties of the detector network in Table \ref{table:detectors}. The CE is proposed to be built in two phases ~\citep{Hall:2020dps}, with the second phase having better sensitivity. We consider the design sensitivity for both detectors. $E_1$, $E_2$, and $E_3$ in our notation denote the three effectively independent detectors that make up the ET observatory. E1, E2, and E3 are located at the corner of the triangle design of ET. We use a 3G detector network of CE-USA + CE-Australia + ET (E1-E2-E3) to do realistic simulations. We choose the 3G detector network to demonstrate the differences solely because signals will likely be present when the detector undergoes non-stationary times. We emphasize that the methods described here will apply to any scenario with long-duration signals where the data can be divided into smaller chunks of stationary (or semi-stationary) times to estimate the \ac{psd} along the signal track.

In the 3G detector network described above, we inject the BNS signal and introduce arbitrary non-stationarities modeled in equation \ref{eqn:noise_model} in the individual detectors as follows:
\begin{itemize}
    \item $CE$-USA: We introduce non-stationarity described by a step function $B(t) = B_0$ for $t > t_c-180$ seconds,i.e., the noise floor is raised around 180 seconds before the trigger time $t_c$. 
    \item $CE$-Australia: Here, we introduce the non-stationarity in the detector noise by dropping the noise floor of the detector at the base level 10 seconds before the trigger time. Mathematically, $B(t) = B_0 $ for $t < t_c-10$.
    \item $E_1$: In this detector we again used the positive step function $B(t)=B_0$ for $t > t_c - 50$ seconds.
    \item $E_2$: Noise floor drops to base scale 50 seconds before trigger time: $B(t)=B_0$ for $t < t_c - 50$.
    \item $E_3$: No non-stationarity is introduced in the $E_3$ detector.
\end{itemize}
\noindent
These injections are different from ones used in single-detector SNR cases in table \ref{table:nonstationary_scenarios}. We use $B_0=0.8$ for all the injections described above.
It is essential to understand that the overall change in fractional SNR across a network of detectors, caused by different periods of non-stationary behavior, will differ from that of an individual detector. Dealing with non-stationary noise in a detector network is more complicated due to the correlation between various parameters and the unique non-stationary characteristics of each detector. However, we can still keep track of the non-stationarity in individual detectors, create a pseudo-\ac{psd} for each detector based on the reference signal, and use it for \ac{pe}.

We do Bayesian \ac{pe} using publicly available code PyCBC Inference~\citep{Biwer:2018osg}. We use uniform priors in detector frame chirp mass $\mathcal{M}$, mass ratio $(q=m_1/m_2); m_1\geq m_2$, comoving volume, and time of arrival $t_c$. We use the isotropic prior for RA, dec, inclination angle, and polarization. We use the TaylorF2 waveform model \citep{Blanchet:1995ez, Faye:2012we} implemented in \textsc{LALSuite} \citep{lalsuite} to simulate the signal to be injected in the noise, and for recovery while doing \ac{pe}. We estimate the likelihood function using the heterodyne likelihood model described in ~\citep{Cornish:2010kf, Finstad:2020sok, Zackay:2018qdy}. We use the sample rate of 1024 Hz to reduce computational costs to generate the data. We use low-frequency cutoff $f_\mathrm{low}=5.1$ Hz for CE detectors and $f_\mathrm{low}=5$Hz for ET, and high-frequency cutoff $f_\mathrm{high}=512$ Hz (for the full \ac{pe} run) for all detectors to estimate likelihood function. A significant contribution of the SNR for a BNS signal comes from the inspiral part. 

Apart from the full \ac{pe} runs, we also do the \ac{pe} runs for scenarios where a pre-merger localization alert can be created. It is done by doing this analysis introducing a cutoff at a high frequency consistent with the time to merge. For example, from the time-frequency track of the signal in figure \ref{fig:signal_track}, the cutoff frequency of $\approx10$ Hz corresponds to 1000 seconds before the merger. To create a pre-merger alert for this signal 1000 seconds before the merger, we must do a fast \ac{pe} with a high-frequency cutoff of less than 10 Hz. To produce pre-merger localization posterior samples, we introduce a cutoff at a high frequency consistent with the time to merge.

\begin{table*}
    \caption{The specifications of the 3G detector network (location and low-frequency cutoff $f_\mathrm{low}$) are considered in this study. We use the design sensitivity noise curves for ET and CE. These detector configurations for CE and ET have also been used in previous work:  \citep{2019CQGra..36v5002H, Nitz2021premerger, Kumar:2021aog}
    }
    \label{table:detectors}
\begin{center}
\begin{tabular}{llllcc}
Observatory  & $f_{\textrm{low}}$ &  Latitude & Longitude \\ \hline
Cosmic Explorer USA & 5.1 & 40.8 & -113.8    \\
Cosmic Explorer Australia & 5.1 &-31.5 & 118.0   \\
Einstein Telescope & 5 & 43.6 & 10.5  \\

\end{tabular}
\end{center}
\end{table*}

\begin{figure*}
    \centering
    \subfigure[]{\includegraphics[width=0.49\textwidth]{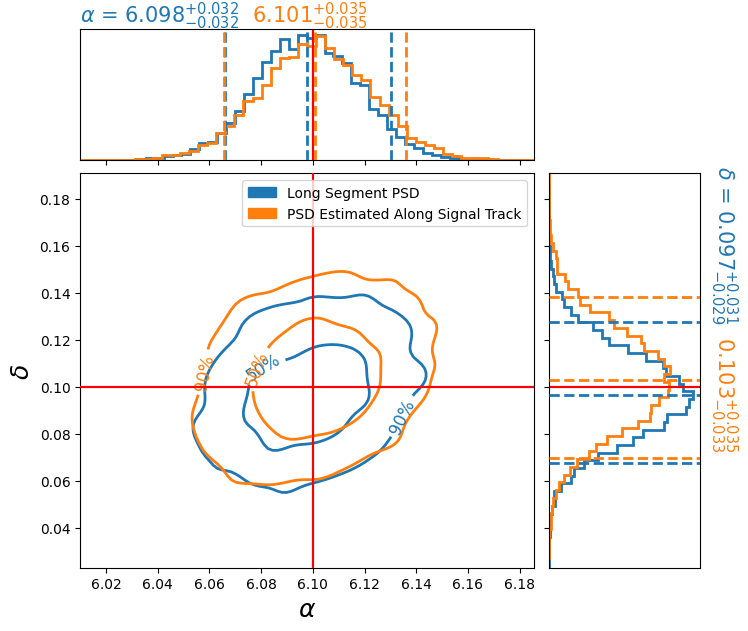}} 
    \subfigure[]{\includegraphics[width=0.49\textwidth]{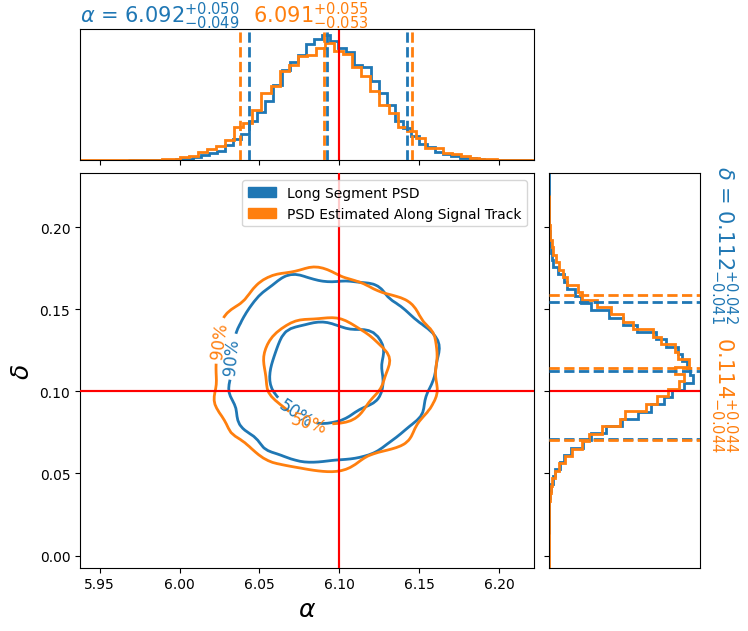}} 
    \caption{This figure shows the difference in localization area for right ascension and declination angles (in radians) arising for two cases. Blue contours represent the $50\%$ and $90\%$ credible intervals when a constant \ac{psd} is used for \ac{pe}. The orange contours represent the same when a \ac{psd} estimated along the track of the signal is used. For this simulation, we used a loud BNS injection ($m_\mathrm{1, source}:1.5~M_\odot$, $m_\mathrm{2, source}:1.3~M_\odot$, $d_L:1000$Mpc) in a third generation detector network of two cosmic explorer and Einstein telescope. Solid red lines show the injected values. The left panel shows the localization area $\sim60$ seconds before the merger, and the right panel shows the same $\sim 180$ seconds before the merger. See the text for details on the detector network and the types of non-stationarity considered.}
    \label{fig:pe_radec}
\end{figure*}

In figure \ref{fig:pe_radec}, we show the difference in the sky localization area inferred from two methods of estimating the \ac{psd}. We do \ac{pe} runs and produce sky area for two pre-merger scenarios: 60 seconds before and 180 seconds before. The difference in network SNR corresponding to maximum likelihood obtained for this example is $~\sim 10\%~(5\%)$ with the corresponding difference in the area in RA-dec plane turns out to be $\sim 20\%~(13\%)$ for \ac{pe} run 60 seconds (180 seconds) before the merger. It shows that there can be significant changes in the reported localization volume if the \ac{psd} corrections are not applied due to non-stationarity. We observe similar trends in the other binary parameters like chirp mass and luminosity distance.

\noindent

\begin{figure}[ht]
\includegraphics[width=0.48\textwidth]{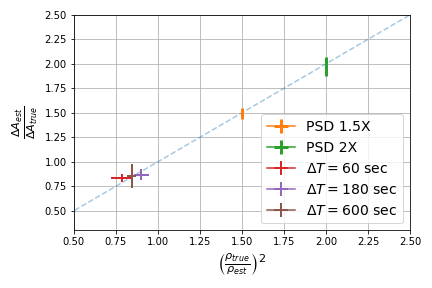}
\caption{Here we show the relation $\Delta A \propto \frac{1}{\rho^2}$ which is represented by a blue dashed line. We use a set of 10 injections for randomly chosen RA-dec values in the detector network described by two CE and one ET for each scenario. For the test purpose, we do two sets of \ac{pe} runs with wrongly estimated \ac{psd} where we scale the estimated \ac{psd} $S_\mathrm{est}(f) = \alpha S_\mathrm{true}(f)$. We set $\alpha=1.5$ and $2.0$ for cases labeled as \ac{psd} 1.5X and 2X, respectively. Other scenarios represent the \ac{pe} done for some time before the merger. $\Delta T = X$ sec means the \ac{pe} for the BNS events X seconds before the merger happens. The error bars in the value represent the spread in the sky area ratio and SNR ratio.}
\label{fig:area_snr_plot}
\end{figure}

From now on, we call the \ac{psd} along the track of signal as `corrected' or `True' \ac{psd} and the one measured using the standard method, without considering non-stationarity as `estimated' \ac{psd}. For two-dimensional localization volumes, e.g., RA-dec posterior samples, the area ($\Delta A$) should scale as the inverse square of the SNR ($\rho$), i.e. $\Delta A\propto 1/\rho^2$. So, a 10 \% fractional error in SNR can induce up to 21 \% fractional errors in 2D localization volume. However, a detector network might depend on various factors as different detectors are sensitive to the different sky areas, and there might be a strong correlation between various signal parameters. To test this scaling, we perform ten injections of the BNS signals in the detector network described above but with random sky localization (keeping all other parameters fixed). We do the \ac{pe} runs with `corrected' and `estimated' \ac{psd}s. In figure \ref{fig:area_snr_plot}, we plot the spread of $\rho_\mathrm{true}/\rho_\mathrm{est}$ vs $\Delta A_\mathrm{est}/\Delta A_\mathrm{true}$. We call $\rho_\mathrm{true}$ the SNR corresponding to the \ac{pe} run with corrected \ac{psd} and $\rho_\mathrm{est}$ the SNR corresponding to the \ac{pe} run with estimated \ac{psd}. Similarly, $\Delta A_\mathrm{true}$ is the 2D area from RA-dec posterior samples for the \ac{pe} run with corrected \ac{psd}. $\Delta A_\mathrm{est}$ is the same with estimated \ac{psd}. The spread in the data points corresponds to the standard deviation of the values on the x-axis and y-axis for all ten injections in each scenario. As a test scenario, we also do \ac{pe} runs for similar injections in Gaussian and stationary noise with misestimated \ac{psd}, as well as true \ac{psd}. We use a \ac{psd} $S_\mathrm{est}(f)=1.5 S_\mathrm{true}(f)$ and $S_\mathrm{est}(f)=2.0 S_\mathrm{true}(f)$ for all the detectors in the network. For these test scenarios, we expect the 2D sky areas to be misestimated by a factor of 1.5 and 2, respectively, shown in figure \ref{fig:area_snr_plot}.

In this study, we demonstrate that inaccurate estimation of the \ac{psd} can lead to incorrect localization volumes. If not correctly accounted for, non-stationarity can potentially cause such misestimation of the \ac{psd}. The \ac{pe} method we present is as fast as any contemporary \ac{pe} method, as additional computation of pseudo-\ac{psd} along the signal track is not computationally expensive compared to a full \ac{pe} run.

\section{Summary and Discussion}

This work proposes a fast method incorporating the non-stationarity, especially for long signals such as BNS. We propose estimating the \ac{psd} along the signal track in the time-frequency space. We assume that the data can be broken into smaller chunks that can be considered stationary or semi-stationary (such as the slowly varying amplitude of the Gaussian noise). We do simulations by injecting a long BNS signal in various scenarios of arbitrary non-stationary noise in a 3G detector and show that there will be a difference in the SNR estimated if these effects are not accounted for. We also show that it will lead to a difference in the estimated localization volume obtained from \ac{pe}.

This method is reliable for the types of non-stationarities where the strain data can be broken into smaller, piecewise segments of semi-stationary data, such as when the amplitude of the Gaussian noise changes slowly in the smaller time bins. It also relies on identifying nearly the right track of the signal in the time-frequency space. In practice, identifying this track should not be problematic; several bootstrapping schemes are practical, including using the initial point estimates for the coalescence time and chirp mass from standard compact-binary searches.  

The method for calculating a corrected \ac{psd} can also be efficiently applied to searches using a simple two-step procedure. In the first stage, an average \ac{psd} is used to calculate an initial SNR estimate and identify potential candidates. Once the initial candidates are identified, the SNR can be re-weighted using the corrected \ac{psd}. It is a natural extension of the dynamic normalization outlined in~\cite{Mozzon2020}.

The proposed method divides the time-frequency track into bins, allowing the data within each bin to be treated as Gaussian and stationary. Each bin must be at least a couple of seconds long to estimate the \ac{psd} within that bin accurately. However, this requirement limits the method's resolving power, meaning that non-stationarities occurring over shorter segments cannot be effectively resolved, which hampers the creation of an accurate \ac{psd} along the signal track. In the follow-up study, we plan to create an effective \ac{psd} using the \ac{psd} variation statistic and baseline \ac{psd}, where the binning scheme is not required. It might help in resolving non-stationarities in shorter segments.

For future work, we plan to modify this method to include more general kinds of non-stationarities using the \ac{psd} variation statistic. We also plan to apply this method to study non-stationarities present in the data from the current detectors by identifying those stretches and doing \ac{pe} by performing BNS injections. One of the natural extensions of this study is to include the uncertainties in \ac{psd} estimation methods along with the non-stationarities and calibration uncertainties to do a full \ac{pe} that incorporates all these effects together. 

\noindent
\acknowledgments
We acknowledge the Max Planck Gesellschaft. We thank the computing team from AEI Hannover for their significant technical support. SK thanks observational relativity and cosmology group members at AEI for their feedback and valuable comments. SK also thanks Aditya Vijaykumar for the feedback on the manuscript. We used the Atlas cluster at AEI Hannover for all the computation. The authors also thank the anonymous referee for their critical input in improving this manuscript.

\bibliography{references}

\end{CJK*}
\end{document}